\documentclass[twocolumn,showpacs,amsmath,amssymb,superscriptaddress,nofootinbib]{revtex4}

\usepackage{graphicx}

\begin{document}
\title{Asymptotic teleportation scheme as a universal programmable quantum processor}
\author{Satoshi Ishizaka}
\affiliation{Nano Electronics Research Laboratories, NEC Corporation,
34 Miyukigaoka, Tsukuba 305-8501, Japan}
\affiliation{INQIE, the University of Tokyo,
4-6-1 Komaba, Meguro-ku, Tokyo 153-8505, Japan}
\author{Tohya Hiroshima}
\affiliation{Nano Electronics Research Laboratories, NEC Corporation,
34 Miyukigaoka, Tsukuba 305-8501, Japan}
\affiliation{
Quantum Computation and Information Project, ERATO-SORST, Japan Science and Technology Agency,\\
Daini Hongo White Building 201, Hongo 5-28-3, Bunkyo-ku, Tokyo 113-0033, Japan}
\date{\today}
%

\begin{abstract}
We consider a scheme of quantum teleportation where a receiver has multiple
($N$) output ports and obtains the teleported state by merely selecting one of
the $N$ ports according to the outcome of the sender's measurement. We
demonstrate that such teleportation is possible by showing an explicit protocol
where $N$ pairs of maximally entangled qubits are employed. The optimal
measurement performed by a sender is the square-root measurement, and a
perfect teleportation fidelity is asymptotically achieved for a large $N$
limit. Such asymptotic teleportation can be utilized as a universal
programmable processor.
\end{abstract}

%
\maketitle
%

Quantum teleportation \cite{Bennett95b} is a technique to transfer an unknown
quantum state from a sender (Alice) to a receiver (Bob) exploiting their
prior shared entangled state. In the standard teleportation scheme, Alice
first performs a joint measurement on the state to be teleported and half
of their entangled state. She then tells the outcome to Bob
via a classical communication channel. To complete the teleportation, Bob
applies a unitary transformation, which depends on the outcome of the Alice's
measurement, to the remaining half of their entangled state.

On the other hand, programmable processors (in short, processors)
\cite{Nielsen97a,Processor}
are devices to manipulate a state via {\it program states}. Suppose
that we wish to apply operation $\varepsilon$ to an input state
$|\chi_{\rm in}\rangle$ such that
$|\chi_{\rm in}\rangle\rightarrow\varepsilon(|\chi_{\rm in}\rangle)$.
To do this by using a processor, we first generate the program state
$|\varepsilon\rangle$, in which $\varepsilon$ is stored. A processor
then performs a fixed operation $G$ and accomplishes the desired task such that
$G(|\chi_{\rm in}\rangle\otimes|\varepsilon\rangle)=
\varepsilon(|\chi_{\rm in}\rangle)\otimes|\varepsilon'\rangle$,
just like a general-purpose computer executes a program stored in memory.
In this way, a programmable processor provides the scheme
of storing and retrieving operations.
If a processor can deal with arbitrary $\varepsilon$, it is called
a universal (programmable) processor. It was shown that a
faithful [the output state is exactly $\varepsilon(|\chi_{\rm in}\rangle)$] and
deterministic (with a unit success probability) universal processor cannot be
realized by a finite dimensional system \cite{Nielsen97a}. The standard
teleportation scheme provides a probabilistic universal
processor \cite{Nielsen97a}, but the success probability becomes extremely
small if the dimension of an input state is large;
the obstacle is that Bob's unitary transformation in the teleportation scheme
generally does not commute with $\varepsilon$ \cite{Nielsen97a,Brukner03a}.

\begin{figure}[b]
\centerline{\scalebox{0.43}[0.43]{\includegraphics{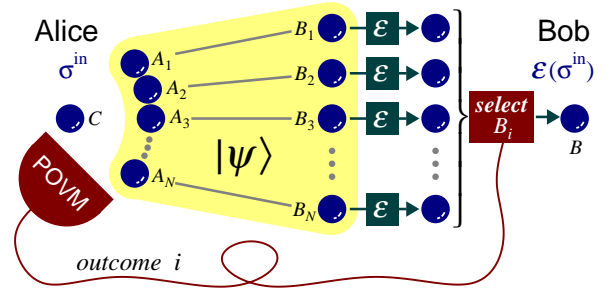}}}
\caption{Setting of asymptotic teleportation.}
\label{fig: Setting}
\end{figure}

Let us then consider the teleportation scheme proposed by Knill, Laflamme, and
Milburn (KLM) \cite{KLM} (and its deterministic version \cite{Franson02a}),
which is a technique to enable linear-optics quantum computation. In the KLM
scheme, Bob has multiple ($N$) output ports and obtains the teleported state by
selecting one of the $N$ ports according to the outcome of Alice's
measurement (see Fig.\ \ref{fig: Setting}). To complete the teleportation,
however, Bob further needs to apply a unitary transformation (phase shift) to
the state of the selected port, as well as the standard teleportation scheme.
If the KLM scheme is successfully modified such that the unitary transformation
is unnecessary (i.e., the state of one of the $N$ ports becomes the teleported
state as it is), the teleportation scheme can provide a universal processor.
Suppose that Bob applies $\varepsilon$ to every port (denoted by
$\varepsilon^{\otimes N}$; see Fig.\ \ref{fig: Setting}) in advance of the
teleportation (this corresponds to the operation for storing $\varepsilon$).
The teleportation procedure then results in the state
processed by $\varepsilon$, regardless of which port is selected. This is
because the operation of selecting a port (without any additional unitary
transformation) always commutes with $\varepsilon^{\otimes N}$, i.e., selecting
a port after applying $\varepsilon^{\otimes N}$ causes the same result as
applying $\varepsilon^{\otimes N}$ after selecting a port. This implies that
the fixed operation of the teleportation (Alice's measurement and Bob's
selection) can {\it execute} arbitrary $\varepsilon$ (including measurements
and even trace-nonpreserving operations), if the state $|\psi\rangle$ employed
for the teleportation is changed into
$|\varepsilon\rangle=(\openone\otimes\varepsilon^{\otimes N})|\psi\rangle$.
Note that, since the form of the program state $|\varepsilon\rangle$ is known
for given $\varepsilon$, we can also generate it by various methods other than
applying $\varepsilon^{\otimes N}$ \cite{Universal}.
This teleportation scheme does
not contradict the law of physics that prohibits superluminal (faster than
light) communication because without knowing the outcome of Alice's
measurement, Bob cannot know which port contains the teleported state and,
hence, cannot obtain any information about the teleported state. However, such
a scheme must be an approximate one if $N$ is finite; otherwise
a faithful and deterministic universal processor would be realized by a finite
dimensional system, which contradicts the no-go theorem in \cite{Nielsen97a}.
Therefore, it is quite desirable to achieve faithful teleportation in the
asymptotic limit of $N\rightarrow\infty$.

In this paper, we demonstrate that such asymptotic teleportation is possible
by showing an explicit protocol where $N$ pairs of maximally entangled qubits
(quantum bits) are employed. A perfect teleportation fidelity is achieved in
the asymptotic limit. Moreover, we determine the optimal measurement performed
by Alice.

Now, let us formulate asymptotic teleportation that aims to
teleport an unknown state on a qudit ($d$-dimensional system). To begin with,
Alice and Bob share a pure entangled state $|\psi\rangle$ on $2N$ qudits
(Fig.\ \ref{fig: Setting}). Bob has half of the $2N$ qudits: $B_1$, $B_2$,
$\cdots$, and $B_N$, where each corresponds to the output port, i.e.,
the unknown state of Alice's $C$ qudit is finally teleported to one of the $N$
qudits. Alice has the remaining half of the $2N$ qudits: $A_1$, $A_2$,
$\cdots$, and $A_N$. These $N$ qudits are denoted by $A$ as a whole. Without
loss of generality, $|\psi\rangle$ can be written as
\begin{equation*}
|\psi\rangle=(O_A\otimes\openone_{B_1\cdots B_N})
|\phi^+\rangle_{A_1B_1}|\phi^+\rangle_{A_2B_2} \cdots
|\phi^+\rangle_{A_NB_N},
\end{equation*}
where $|\phi^+\rangle=(1/\sqrt{d})\sum_{k=0}^{d-1} |kk\rangle$, and $O$ is an
arbitrary operator that satisfies $\hbox{tr}OO^\dagger=d^N$ so that
$|\psi\rangle$ is normalized. Alice then performs a joint measurement with $N$
possible outcomes ($1$, $2$, $\cdots$, $N$) on the $A$ and $C$ qudits. The
measurement is described by a positive operator valued measure (POVM) whose
elements are $\{\Pi_i\}$ such that $\sum_{i=1}^{N}\Pi_i=\openone_{AC}$. Suppose
that she obtains the outcome $i$. She then tells the outcome to Bob via a
classical channel. Finally, Bob discards the $(N-1)$ qudits of
$B_1B_2\cdots B_{i-1}B_{i+1}\cdots B_N$ (i.e., all his qudits except for
$B_i$), which are briefly denoted by ${\bar B}_i$. The state of the remaining
$B_i$ qudit (regarded as $B$) is the teleported state.

The channel of the above asymptotic teleportation, which maps the density
matrices acting on the Hilbert space ${\cal H}_C$ to those on ${\cal H}_B$,
is thus
\begin{align*}
\Lambda(\sigma^{\rm in})
&=
\sum_{i=1}^{N}\left[\hbox{tr}_{A{\bar B_i}C}
\sqrt{\Pi_i}(|\psi\rangle\langle\psi|\otimes\sigma^{\rm in}_{C})
\sqrt{\Pi_i}^\dagger\right]_{B_i\rightarrow B} \cr
&=
\sum_{i=1}^{N}\hbox{tr}_{AC}
\Pi_i \left(
[(O\otimes\openone)\sigma^{(i)}_{AB}(O^\dagger\otimes\openone)]
\otimes \sigma^{\rm in}_{C}
\right)
\end{align*}
with
\begin{align}
\sigma^{(i)}_{AB}&=\left[\hbox{tr}_{\bar B_i}(
P^+_{A_1B_1}\otimes P^+_{A_2B_2}\otimes \cdots \otimes P^+_{A_NB_N}
)\right]_{B_i\rightarrow B} \nonumber \\
&=\frac{1}{d^{N-1}}P^+_{A_iB} \otimes \openone_{\bar A_i},
\label{eq: sigma}
\end{align}
where $P^+=|\phi^+\rangle\langle\phi^+|$, and ${\bar A}_i$ is a shorthand
notation
for $A_1A_2\cdots A_{i-1}A_{i+1}\cdots A_N$. The channel is characterized by
the fidelity $f$ averaged over all uniformly distributed input pure states,
which is given by $f=(Fd+1)/(d+1)$ with $F$ being the entanglement fidelity of
the channel \cite{Horodecki99a}. For the channel $\Lambda$, we have
\begin{align}
F
&=
\hbox{tr}P^+_{BD}\left[(\Lambda\otimes\openone)P^+_{CD}\right] \nonumber \\
&=
\hbox{tr}\sum_{i=1}^{N} P^+_{BD}
\Pi_{iAC} \left(
[(O\otimes\openone)\sigma^{(i)}_{AB}(O^\dagger\otimes\openone)]
\otimes P^+_{CD}\right) \nonumber \\
&=
\frac{1}{d^2}
\sum_{i=1}^{N} \hbox{tr}\Pi_{iAB}
[(O\otimes\openone)\sigma^{(i)}_{AB}(O^\dagger\otimes\openone)].
\label{eq: F}
\end{align}
Note that $\Pi_i$ is changed into an operator acting on 
${\cal H}_A\otimes{\cal H}_B$ in the last equality of Eq.\ (\ref{eq: F})
because we used the relationship that
$(V\otimes\openone)P^+=(\openone\otimes V^T)P^+$ for any operator $V$. 
Hereafter, the subscript of $AB$ in both $\Pi_i$ and $\sigma^{(i)}$ is omitted
for simplicity, unless it is confusing.

Let us first consider the important case where $O=\openone$ and $d=2$, i.e.,
$N$ pairs of maximally entangled qubits are employed for asymptotic
teleportation. The entanglement
fidelity $F$ in this case is $F=(1/4)\sum_{i=1}^N \hbox{tr}\Pi_i\sigma^{(i)}$.
Therefore, the problem of maximizing $F$ with respect to $\{\Pi_i\}$ is
equivalent to the quantum detection problem of minimizing the error probability
($p_e=1-4F/N$) of the quantum signals
$\{\sigma^{(1)},\sigma^{(2)},\cdots,\sigma^{(N)}\}$ with equal prior
probability $1/N$. The signal states, $\sigma^{(i)}$'s, given by
Eq.\ (\ref{eq: sigma}) are mutually non-commutable mixed states, and therefore
determining the optimal detection measurement in an analytical way is not easy.
Fortunately, however, it can be shown that the square-root measurement (SRM)
(also known as a pretty good measurement or least-squares measurement)
\cite{SRM,Eldar04a} 
is indeed optimal for $\{\sigma^{(i)}\}$.

The POVM elements of SRM are given by
\begin{equation*}
\Pi^{\rm SQ}_i=\rho^{-\frac{1}{2}}\sigma^{(i)}
\rho^{-\frac{1}{2}} \hbox{~~~~with~~~~}
\rho=\sum_{i=1}^{N}\sigma^{(i)}.
\end{equation*}
Since $\rho$ is not full rank, $\rho^{-1}$ is defined on the support of $\rho$.
Moreover, we implicitly assume that 
$\Delta=(\openone-\sum_{i=1}^N \Pi^{\rm SQ}_i)/N$ is added to every
$\Pi^{\rm SQ}_i$ so that the POVM elements sum to identity. Note that the
excess term $\Delta$ does not affect the entanglement fidelity because
$\hbox{tr}\sigma^{(i)}\Delta=0$.

Based on the obvious correspondence between qubits and 1/2 spins, 
$|0(1)\rangle\leftrightarrow|\frac{1}{2},-\frac{1}{2}(\frac{1}{2})\rangle$,
we regard each qubit as a 1/2 spin, i.e., $SU(2)$ basis. It is then convenient
to consider $|\psi\rangle$ of $N$ pairs of spin singlets, i.e.,
$|\psi\rangle=|\psi^-\rangle^{\otimes N}$ (instead of
$|\psi\rangle=|\phi^+\rangle^{\otimes N}$), and as a result $P^+$ in
$\sigma^{(i)}$ is replaced by $P^-=|\psi^-\rangle\langle\psi^-|$ where
$|\psi^-\rangle=(|01\rangle-|10\rangle)/\sqrt{2}$. The POVM elements for the
two cases are easily interconverted by applying the unitary transformation
$(\openone_A\otimes\sigma_{yB})$. In the language of $SU(2)$ representation,
eigenvectors with eigenvalues $\lambda_{j}^{-}=(N/2-j)/2^{N}$ and 
$\lambda_{j}^{+}=(N/2+j+1)/2^{N}$ of $\rho$ are given by 
\begin{align*}
| \Psi_{\mp}^{[N]}(\lambda_{j}^{\mp};m,\alpha) \rangle 
=&\pm | 0 \rangle_{B} | \Phi^{[N]}(j,m_{+},\alpha) \rangle
 \langle j,m_{+};j_{\pm} \rangle_{-} \nonumber \\
 &\pm | 1 \rangle_{B} | \Phi^{[N]}(j,m_{-},\alpha) \rangle 
 \langle j,m_{-};j_{\pm} \rangle_{+}, 
\end{align*}
where $| \Phi^{[N]}(j,m,\alpha) \rangle = | j,m,\alpha \rangle$ denotes the
orthogonal basis of $N$-spin systems, i.e., the basis of irreducible
representation of $SU(2)^{\otimes N}$. The spin angular momentum $j$ runs from
$j_{\rm min}$ to $N/2$ ($m=-j,\dots,j$), where $j_{\rm min}=0$ $(1/2)$ when $N$
is even (odd), and $\alpha$ specifies the additional degree of freedom. Here,
we introduced a shorthand notation for (nonvanishing) Clebsch-Gordan 
coefficients, 
\begin{align*}
\langle j_{1},m_{1};j \rangle_{\pm}
=&\textstyle\langle j_{1},m_{1},\frac{1}{2},\pm \frac{1}{2} | j,m_{1}\pm \frac{1}{2} \rangle  \\
=&\textstyle(-1)^{j_{1}+\frac{1}{2}-j} \langle \frac{1}{2},\pm \frac{1}{2},j_{1},m_{1}| j,m_{1}\pm \frac{1}{2} \rangle,
\end{align*}
and write $j_{\pm}=j\pm \frac{1}{2}$ and $m_{\pm}=m\pm \frac{1}{2}$. The
proof of the eigenvalue equation 
\begin{equation} \label{eq:eigenvalue_equation}
\rho| \Psi_{\mp}^{[N]}(\lambda_{j}^{\mp};m) \rangle =
\lambda_{j}^{\mp}| \Psi_{\mp}^{[N]}(\lambda_{j}^{\mp};m) \rangle .
\end{equation}
is carried out by induction and by noting that $\rho=\rho^{[N]}$ is constructed
recursively:
\begin{equation*}
\rho^{[N]}=\rho^{[N-1]}\otimes \frac{\openone_{A_{N}}}{2}
+\frac{\openone_{A_{1}}}{2} \otimes \dots \otimes \frac{\openone_{A_{N-1}}}{2}
\otimes P_{BA_{N}}^{-}.
\end{equation*}
Details are presented in Appendix A.

The $N$-spin eigenfunctions $|\Phi^{[N]}\rangle$ are computed recursively:
$|\Phi^{[N-1]}\rangle|\Phi^{[1]}\rangle\rightarrow|\Phi^{[N]}\rangle $,
where $|\Phi^{[N-1]}\rangle$ are $(N-1)$-spin eigenfunctions of the first
$(N-1)$ spins ($A_{1},\dots,A_{N-1}$) and $|\Phi^{[1]}\rangle$ are the
1/2 spin function of the $A_N$ qubit. The other choice of construction of
$|\Phi^{[N]}\rangle$ results in a different set of functions,
$|\Phi^{[N]\prime}(j,m,\alpha^{\prime})\rangle$, which are unitarily equivalent
to $|\Phi^{[N]}(j,m,\alpha)\rangle$, and the unitary transformation depends
only on $\alpha$ and $\alpha^{\prime}$ for each $j$. This fact enables us to
calculate explicitly the matrix elements involved in the calculation of the
entanglement fidelity $F$ as follows:
\begin{equation} \label{eq:matrix_element}
\langle \xi^{(i)}(s,s_z,\alpha) | \rho^{-\frac{1}{2}}| \xi^{(i)}(s^{\prime},s_z^{\prime},\alpha^{\prime}) \rangle
= \delta_{s,s^{\prime}} \delta_{s_z,s_z^{\prime}} \delta_{\alpha,\alpha^{\prime}}c(s)
\end{equation}
with
\begin{equation} \label{eq:c}
c(s)=\left( \lambda_{s-\frac{1}{2}}^{-}\right)^{-\frac{1}{2}} \frac{s}{2s+1}
    +\left( \lambda_{s+\frac{1}{2}}^{+}\right)^{-\frac{1}{2}} \frac{s+1}{2s+1}.
\end{equation}
Here, the states of
\begin{equation*}
| \xi^{(i)}(s,s_z,\alpha) \rangle =
| \psi^{-}\rangle_{BA_{i}}
| \Phi^{[N-1]^{\prime}}(s,s_z,\alpha) \rangle
\end{equation*}
with $| \Phi^{[N-1]^{\prime}} \rangle $ being the $(N-1)$-spin eigenfunction
for the ${\bar A}_i$ qubits, are the eigenfunctions of $\sigma^{(i)}$,
and thus,
\begin{equation*}
\sigma^{(i)}=
\frac{1}{2^{N-1}} \sum_{s=s_{\rm min}}^{(N-1)/2}\sum_{s_{z},\alpha} | \xi^{(i)}(s,s_{z},\alpha) \rangle
\langle \xi^{(i)}(s,s_{z},\alpha) |,
\end{equation*}
where $s_{\rm min}=0$ $(1/2)$ when $N-1$ is even (odd). Note that matrix
elements of Eq.~(\ref{eq:matrix_element}) depend only on $s$. See Appendix B
for the detailed derivations of Eqs.~(\ref{eq:matrix_element}) and (\ref{eq:c}). Note further
that, in terms of $|\xi^{(i)}(s,s_z,\alpha)\rangle$, both $\rho$ and
$\sigma^{(i)}$ are found to be block-diagonal with respect to $s$. The block
matrices are denoted by $\rho(s)$ and $\sigma^{(i)}(s)$, respectively.

\begin{figure}[t]
\centerline{\scalebox{0.43}[0.43]{\includegraphics{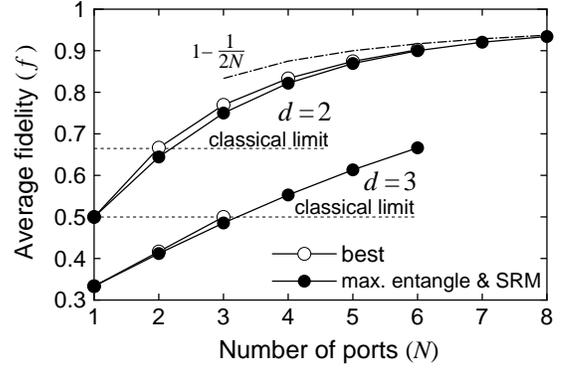}}}
\caption{Average fidelity ($f$) of asymptotic teleportation as 
function of number of output ports ($N$).}
\label{fig: fidelity}
\end{figure}

Now, from Eqs.\ (\ref{eq:matrix_element}) and (\ref{eq:c}), we have 
\begin{align*}
F&=\frac{1}{2^2}\sum_{s=s_{\min }}^{(N-1)/2} \sum_{i=1}^N 
\hbox{tr}\rho(s)^{-\frac{1}{2}} \sigma^{(i)}(s) \rho(s)^{-\frac{1}{2}} \sigma^{(i)}(s) \cr
&=\frac{N}{2^{2N}}\sum_{s=s_{\min }}^{(N-1)/2} \sum_{s_z,\alpha}c(s)^2 \cr
&=\frac{N}{2^{2N}}\sum_{s=s_{\min }}^{(N-1)/2}
\frac{(2s+1)^2 (N-1)!}{(\frac{N-1}{2}-s)!(\frac{N+1}{2}+s)!} c(s)^2 \cr
&=\frac{1}{2^{N+3}}\sum_{k=0}^{N}
\bigg(\frac{N-2k-1}{\sqrt{k+1}}+\frac{N-2k+1}{\sqrt{N-k+1}}\bigg)^2
\binom{N}{k}.
\end{align*}
Here, we introduced $k=(N-1)/2-s$ in the last equality. The corresponding
average fidelity $f$ as a function of $N$ is plotted by closed circles in
Fig.\ \ref{fig: fidelity}. For $N\ge3$, the fidelity exceeds the classical
limit $f_{\rm cl}=2/(d+1)$ ($f_{\rm cl}=2/3$ for $d=2$), which is the
best fidelity via a classical channel only \cite{Horodecki99a}.
Therefore, this protocol works
as quantum teleportation for $N\ge3$. Moreover, the fidelity approaches to
$f=1$ for increasing $N$. In fact, by expanding $(1-x)^{-1/2}$ in $F$ into the
Taylor series of $\frac{(N-2k)^2}{(N+2)^2}$ and noting
$y(m)\equiv\sum_{k=0}^{N}(N-2k)^{2m}\binom{N}{k}
={\cal O}(N^m)2^N
=[(x\frac{d}{dx})^{2m}(x+\frac{1}{x})^N]_{x=1}$
[with $y(1)=N2^N$ and $y(2)=N(3N-2)2^N$], we find that $f\rightarrow1-1/(2N)$
for $N\rightarrow\infty$. Therefore, the protocol of employing $N$
spin singlets and SRM certainly achieves perfect fidelity in the asymptotic
limit.

Let us then prove that SRM is an optimal measurement for $|\psi\rangle$ of
$N$ spin singlets. The problem of maximizing
$F=(1/4)\sum_{i=1}^N \hbox{tr}\Pi_i\sigma^{(i)}$ is a semidefinite program
\cite{Boyd04a} and thus has the dual problem of minimizing
$(1/4)\hbox{tr}Y$ subject to $Y-\sigma^{(i)}\ge0$ for all $i$
\cite{Eldar03a,Eldar04a}. Any feasible solution of the dual problem gives an
upper bound of the original problem. Therefore, it is enough to show that
$Y^{\rm SQ}=\sum_{i=1}^N\Pi^{\rm SQ}_i\sigma^{(i)}$ is a feasible solution
(i.e., $Y^{\rm SQ}-\sigma^{(i)}\ge0$) because $(1/4)\hbox{tr}Y^{\rm SQ}$
agrees with $F$ obtained by SRM. Using Eq.\ (\ref{eq:matrix_element}),
we find that 
$Y^{\rm SQ}=\sum_{s=s_{\rm min}}^{(N-1)/2} Y^{\rm SQ}(s)$ with
$Y^{\rm SQ}(s)=\frac{c(s)}{2^{N-1}}\rho(s)^{\frac{1}{2}}$.
It has been shown that
$A-(1/c)|\xi\rangle\langle\xi|\ge0$ if $|\xi\rangle\in\hbox{range}(A)$ and
$c=\langle \xi|A^{-1}|\xi\rangle$ \cite{Lewenstein98a}.
Moreover, 
$A-(1/c)(|\xi\rangle\langle\xi|+|\xi_{\perp}\rangle\langle\xi_{\perp}|)\ge0$ 
for $|\xi_{\perp}\rangle$ such that $\langle\xi_{\perp}|\xi\rangle=0$,
$|\xi_{\perp}\rangle\in\hbox{range}(A)$, and
$c=\langle \xi_{\perp}|A^{-1}|\xi_{\perp}\rangle$ because
$\langle \xi_{\perp}|[A- (1/c)|\xi\rangle\langle \xi|]^{-1}|\xi_{\perp}\rangle
=c$ \cite{Lewenstein98a}. Repeating this, it is found that
$A-(1/c)\sum_k |\xi_k\rangle\langle\xi_k|\ge0$ where $|\xi_k\rangle$ are
mutually orthogonal vectors such that $|\xi_k\rangle\in\hbox{range}(A)$ and
$\langle\xi_k|A^{-1}|\xi_k\rangle=c$. Therefore, 
\begin{equation*}
\rho(s)^{\frac{1}{2}}-\frac{1}{c(s)}\sum_{s_z,\alpha}|\xi^{(i)}(s,s_z,\alpha)\rangle\langle\xi^{(i)}(s,s_z,\alpha)|\ge0
\end{equation*}
follows from Eq.\ (\ref{eq:matrix_element}), and thus
$Y^{\rm SQ}(s)-\sigma^{(i)}(s)\ge0$, which completes the proof of the
optimality.

Let us return to Eq.\ (\ref{eq: F}) and investigate the cases of general $O$.
We need to optimize both $\{\Pi_i\}$ and $O$ to obtain the best fidelity of
asymptotic teleportation. By introducing
$\tilde \Pi_i=(O^\dagger\otimes \openone) \Pi_i (O\otimes \openone)$
and $X=O^\dagger O$, however, the best entanglement fidelity is obtained by
maximizing
\begin{equation}
F=\frac{1}{d^2}\sum_{i=1}^N \hbox{tr}\tilde \Pi_{iAB} \sigma^{(i)}_{AB}
\label{eq: Primary problem}
\end{equation}
under the constraints of $\tilde\Pi_i\ge0$,
$\sum_{i=1}^N\tilde\Pi_i=(X\otimes \openone)$, $X\ge0$, and $\hbox{tr}X=d^N$.
This is also a semidefinite program.
The dual problem is of
minimizing $F=d^{N-2}a$ subject to $\Omega-\sigma^{(i)}\ge0$ and
$a\openone-\hbox{tr}_B \Omega\ge0$. The constraints are satisfied if we take
$\Omega=\sum_{i=1}^N \sigma^{(i)}$ and $a=N/d^N$, and hence we have
$F\le N/d^2$. This upper bound is tight for $N\le d$. In fact, letting
$\{|e_k\rangle\}$ be the set of $N$ orthogonal states on ${\mathbb C}^d$, the
protocol of employing the separable
$|\psi\rangle=\bigotimes_{k=1}^N|0\rangle_{A_k}|e_k\rangle_{B_k}$, which
results in mutually orthogonal
$(O\otimes\openone)\sigma^{(i)}(O^\dagger\otimes\openone)=
(|0\rangle\langle0|)_A\otimes(|e_i\rangle\langle e_i|)_B$,
achieves the upper bound. The corresponding bound for the average fidelity is
$f\le(d+N)/[d(d+1)]\le f_{\rm cl}$, and therefore it is concluded that $N>d$
is necessary for any protocol to exceed the classical limit of fidelity.

For $N>d$, such a construction of $N$ orthogonal states becomes impossible
(even using entangled states), and the best $F$ may deviate from $N/d^2$. We
have solved the semidefinite program Eq.\ (\ref{eq: Primary problem}) using
the numerical package of SDPA \cite{SDPA}. The results for $d=2$ and $N\le6$
are plotted by open circles in Fig.\ \ref{fig: fidelity}. It is found from the
figure that the best fidelity is nearly achieved by the protocol of
employing spin singlets (maximally entangled $|\psi\rangle$) and SRM. 
Interestingly, although the difference is small,
this implies that a non-maximally entangled $|\psi\rangle$
provides a higher fidelity than that of the maximally entangled $|\psi\rangle$.
In Fig.\ \ref{fig: fidelity}, the fidelity for the case of maximally
entangled qutrits ($d=3$) and SRM is also plotted, which was obtained by
the numerical diagonalization of $\rho$. The numerical
investigations suggest that SRM is optimal even in this case.
For the case of maximally entangled qudits with general $d$ and SRM,
it can be shown by the same technique as used in \cite{Holevo98a} that
$f\ge1-d(d-1)/N$. Therefore, the protocol provides a perfect fidelity in the
asymptotic limit for any $d$.

To summarize, we considered a scheme of asymptotic quantum teleportation where
Bob has multiple output ports and obtains the teleported state by simply
selecting one of the $N$ ports. We showed that, if $N$ pairs of maximally
entangled qubits are employed, the square-root measurement is the optimal
measurement performed by Alice. This protocol provides a perfect fidelity in
the asymptotic limit and nearly achieves the best fidelity of asymptotic
teleportation. The scheme of asymptotic teleportation provides a universal
programmable processor in a simple and natural way: to process a state by
operation $\varepsilon$, teleport the state by employing
$|\varepsilon\rangle=(\openone\otimes\varepsilon^{\otimes N})|\psi\rangle$
instead of $|\psi\rangle$. The fidelity of the processor
dealing with trace-preserving
$\varepsilon$ is always equal to or higher than the fidelity of asymptotic
teleportation alone because of the monotonicity
such that $f(\varepsilon(|\chi_{\rm in}\rangle),
\varepsilon(\Lambda(|\chi_{\rm in}\rangle))) 
\ge f(|\chi_{\rm in}\rangle,\Lambda(|\chi_{\rm in}\rangle))$, where $\Lambda$
is the teleportation channel, and therefore an asymptotically faithful
programmable processor is realized.

This work was supported by the Special Coordination Funds for Promoting
Science and Technology.
%

%

%

\begin{widetext}

\appendix

\section{Proof of Eq.~(\ref{eq:eigenvalue_equation})}

In this appendix, we drop $\alpha$ from 
$\left| \Phi^{[N]([N-1])} \right\rangle $ and $\left| \Psi^{[N]([N-1])} \right\rangle $ for simplicity.
We note that $\left| \Phi^{[N]}(j,\dots) \right\rangle $ is classified into two;
one is the linear combination of 
$\left| \Phi^{[N-1]}(j_{+},\dots) \right\rangle \left| i \right\rangle_{A_{N}}$ 
and the other, $\left| \Phi^{[N-1]}(j_{-},\dots) \right\rangle \left|i \right\rangle_{A_{N}}$.
We call the former (latter) is of the type-I (II);
\begin{equation*}
\left| \Phi_{\mathrm{I}(\mathrm{II})}^{[N]}(j,m) \right\rangle
=\left| \Phi^{[N-1]}(j_{+}(j_{-}),m_{+}) \right\rangle 
  \left| 0\right\rangle_{A_{N}} \left\langle j_{+}(j_{-}),m_{+};j \right\rangle_{-} 
+\left| \Phi^{[N-1]}(j_{+}(j_{-}),m_{-})\right\rangle 
  \left| 1\right\rangle_{A_{N}} \left\langle j_{+}(j_{-}),m_{-};j \right\rangle_{+} .
\end{equation*}
According to the different types of $\left| \Phi^{[N]} \right\rangle $, 
$\left| \Psi_{\mp}^{[N]} \right\rangle $ have two types; 
\begin{align} \label{eq:Psi_1}
\left| \Psi_{\mp \mathrm{I}(\mathrm{II})}^{[N]}(\lambda_{j}^{\mp},m) \right\rangle 
=&\pm    \left| 0 \right\rangle_{B} \left| \Phi^{[N-1]}(j_{+}(j_{-}),m_{++}) \right\rangle
         \left| 0 \right\rangle_{A_{N}} 
  \left\langle j,m_{+};j_{\pm} \right\rangle_{-}
         \left\langle j_{+}(j_{-}),m_{++};j\right\rangle_{-} \nonumber \\
 &\pm    \left| 0 \right\rangle_{B} \left| \Phi^{[N-1]}(j_{+}(j_{-}),m) \right\rangle
         \left| 1 \right\rangle_{A_{N}} 
  \left\langle j,m_{+};j_{\pm}\right\rangle_{-}
         \left\langle j_{+}(j_{-}),m;j \right\rangle_{+} \nonumber \\
 &\pm    \left| 1 \right\rangle_{B} \left| \Phi^{[N-1]}(j_{+}(j_{-}),m) \right\rangle
         \left| 1 \right\rangle_{A_{N}} 
  \left\langle j,m_{-};j_{\pm} \right\rangle_{+}
         \left\langle j_{+}(j_{-}),m;j \right\rangle_{-} \nonumber \\
 &\pm    \left| 1 \right\rangle_{B} \left| \Phi^{[N-1]}(j_{+}(j_{-}),m_{--}) \right\rangle
         \left| 1 \right\rangle_{A_{N}}
  \left\langle j,m_{-};j_{\pm} \right\rangle_{+}
         \left\langle j_{+}(j_{-}),m_{--};j \right\rangle_{+} ,
\end{align}
where $j_{\pm \pm}=j\pm 1$ and $m_{\pm \pm}=m\pm 1$.
Since Eq.~(\ref{eq:eigenvalue_equation}) is obvious for $N=1$, our aim is 
to prove Eq.~(\ref{eq:eigenvalue_equation})
under the assumption that Eq.~(\ref{eq:eigenvalue_equation}) with $N\rightarrow N-1$ holds true.
To this end, we write 
$\left| \Psi^{[N]} \right\rangle $ in terms of $\left| \Psi^{[N-1]} \right\rangle $ as follows.
\begin{align} \label{eq:Psi_2}
&\hspace*{-0.3cm}\left| \Psi_{\mp \mathrm{I}(\mathrm{II})}^{[N]}(\lambda_{j}^{\mp};m) \right\rangle=  \nonumber \\
&\pm \left| \Psi_{-}^{[N-1]}(\lambda_{j_{+}(j_{-})}^{-};m_{+}) \right\rangle
 \left| 0 \right\rangle_{A_{N}} \nonumber \\
&\times \left[
 \left\langle j_{+}(j_{-}),m_{++};j_{++}(j) \right\rangle_{-}^{*}
 \left\langle j,m_{+};j_{\pm} \right\rangle_{-}
 \left\langle j_{+}(j_{-}),m_{++};j \right\rangle_{-}
+\left\langle j_{+}(j_{-}),m;j_{++}(j) \right\rangle_{+}^{*}
 \left\langle j,m_{-};j_{\pm} \right\rangle_{+}
 \left\langle j_{+}(j_{-}),m;j \right\rangle_{-} 
        \right]  \nonumber \\
&\pm \left| \Psi_{-}^{[N-1]}(\lambda_{j_{+}(j_{-})}^{-};m_{-}) \right\rangle
 \left| 1\right\rangle_{A_{N}} \nonumber \\
&\times \left[
 \left\langle j_{+}(j_{-}),m;j_{++}(j) \right\rangle_{-}^{*}
 \left\langle j,m_{+};j_{\pm} \right\rangle_{-}
 \left\langle j_{+}(j_{-}),m;j \right\rangle_{+}
+\left\langle j_{+}(j_{-}),m_{--};j_{++}(j) \right\rangle_{+}^{*}
 \left\langle j,m_{-};j_{\pm} \right\rangle_{+}
 \left\langle j_{+}(j_{-}),m_{--};j \right\rangle_{+} 
        \right] \nonumber \\
&\mp \left| \Psi_{+}^{[N-1]}(\lambda_{j_{+}(j_{-})}^{+};m_{+}) \right\rangle
 \left| 0 \right\rangle_{A_{N}} \nonumber \\
&\times \left[ 
 \left\langle j_{+}(j_{-}),m_{++};j(j_{--}) \right\rangle_{-}^{*}
 \left\langle j,m_{+};j_{\pm} \right\rangle_{-}
 \left\langle j_{+}(j_{-}),m_{++};j \right\rangle_{-}
+\left\langle j_{+}(j_{-}),m;j(j_{--}) \right\rangle_{+}^{*}
 \left\langle j,m_{-};j_{\pm} \right\rangle_{+}
 \left\langle j_{+}(j_{-}),m;j \right\rangle_{-} \right] \nonumber \\
&\mp \left| \Psi_{+}^{[N-1]}(\lambda_{j_{+}(j_{-})}^{+};m_{-}) \right\rangle
 \left| 1 \right\rangle_{A_{N}} \nonumber \\
&\times \left[
 \left\langle j_{+}(j_{-}),m;j(j_{--}) \right\rangle_{-}^{*}
 \left\langle j,m_{+};j_{\pm} \right\rangle_{-}
 \left\langle j_{+}(j_{-}),m;j \right\rangle_{+}
+\left\langle j_{+}(j_{-}),m_{--};j(j_{--}) \right\rangle_{+}^{*}
 \left\langle j,m_{-};j_{\pm} \right\rangle_{+}
 \left\langle j_{+}(j_{-}),m_{--};j \right\rangle_{+} 
        \right] .
\end{align}
Equation~(\ref{eq:Psi_2}) is obtained by computing the overlap between 
$\left| \Psi_{\mp \mathrm{I}(\mathrm{II})}^{[N]} \right\rangle $ 
given by Eq.~(\ref{eq:Psi_1}) and 
$\left| \Psi_{\mp}^{[N-1]} \right\rangle $ given by Eq.~(\ref{eq:eigenvector}) with $N \rightarrow N-1$. 
The vector 
$\rho^{[N-1]} \otimes I_{A_{N}}
\left| \Psi_{\mp \mathrm{I}(\mathrm{II})}^{[N]}(\lambda_{j}^{\mp};m) \right\rangle $ 
takes the form of the right hand side (r.h.s.) of Eq.~(\ref{eq:Psi_2}) with
\begin{equation} \label{eq:replace}
\left| \Psi_{\mp}^{[N-1]}(\lambda_{j_{+}(j_{-})}^{\mp};\dots) \right\rangle 
\rightarrow
\lambda_{j_{+}(j_{-})}^{[N-1]\mp}
\left| \Psi_{\mp}^{[N-1]}(\lambda_{j_{+}(j_{-})}^{[N-1]\mp};\dots) \right\rangle ,
\end{equation}
while the vector 
$I_{A_{1}} \otimes \dots \otimes I_{A_{N-1}} \otimes P_{BA_{N}}^{-}
\left| \Psi_{\mp \mathrm{I}(\mathrm{II})}^{[N]}(\lambda_{j}^{\mp};m) \right\rangle $ 
takes the form of the r.h.s. of Eq.~(\ref{eq:Psi_1}) with
\begin{align*}
\left| 0 \right\rangle_{B}\left| 1 \right\rangle_{A_{N}} &\rightarrow
\left(
 \left| 0 \right\rangle_{B} \left| 1 \right\rangle_{A_{N}}
-\left| 1 \right\rangle_{B} \left| 0 \right\rangle_{A_{N}} \right) /\sqrt{2} \\
 \left| 1 \right\rangle_{B} \left| 0 \right\rangle_{A_{N}} &\rightarrow
-\left(
 \left| 0 \right\rangle_{B} \left| 1 \right\rangle_{A_{N}}
-\left| 1 \right\rangle_{B} \left| 0 \right\rangle_{A_{N}} \right) /\sqrt{2}.
\end{align*}
In Eq.~(\ref{eq:replace}), we attached a superscript $[N-1]$ to eigenvalues $\lambda^{\mp}$ 
emphasizing the relevant system size. 
Putting these two results together (and after lengthy calculations), 
we can see the desired eigenvalue equation, 
\begin{equation*}
\rho^{[N]} \left| \Psi_{\mp \mathrm{I}(\mathrm{II})}^{[N]}(\lambda_{j}^{\mp};m) \right\rangle =
\lambda_{j}^{\mp}\left| \Psi_{\mp \mathrm{I}(\mathrm{II})}^{[N]}(\lambda_{j}^{\mp};m) \right\rangle .
\end{equation*}
This completes the proof.

\section{Eq.~(\ref{eq:matrix_element}) with Eq.~(\ref{eq:c})}

To calculate the entanglement fidelity $F$, we decompose $\rho^{[N]}$ into operators
acting on eigenspace with eigenvalues $\lambda_{j}^{\mp}$,
\begin{align} \label{eq:decomposition}
\rho_{\mp}^{[N]}(\lambda_{j}^{\mp})= 
&\lambda_{j}^{\mp} \sum_{m=-(j \pm 1/2)}^{j \pm 1/2}
\sum_{\alpha} \left| \Psi_{\mp}^{[N]}(\lambda_{j}^{\mp},m,\alpha) \right\rangle
\left\langle \Psi_{\mp}^{[N]}(\lambda_{j}^{\mp},m,\alpha) \right| ,
\end{align}
so that we can write 
$\rho^{[N]}=\rho_{-}^{[N]} \oplus \rho_{+}^{[N]}$, 
where 
$\rho_{\mp}^{[N]}=\oplus_{j} \rho_{\mp}^{[N]}(\lambda_{j}^{\mp})$.
Let us recall that $\left| \Phi^{[N-1]}(s,s_{z},\alpha) \right\rangle $ is
unitarily equivalent to $\left| \Phi^{[N-1]^{\prime}}(s,s_{z},\alpha^{\prime}) \right\rangle $ 
and the unitary transform is independent of $s_{z}$;
\begin{equation*}
\left| \Phi^{[N-1]}(s,s_{z},\alpha) \right\rangle =
\sum_{\alpha\alpha^{\prime}} U_{\alpha\alpha^{\prime}}(s)
\left| \Phi^{[N-1]^{\prime}}(s,s_{z},\alpha^{\prime}) \right\rangle 
\end{equation*}
to obtain
\begin{align} \label{eq:inner_product}
&\hspace*{-2cm}\left\langle \xi^{(l)}(s,s_{z},\beta) \right|
\left. \Psi_{\mp \mathrm{I}(\mathrm{II})}^{[N]}(\lambda_{j}^{\mp},m,\alpha) \right\rangle 
\nonumber \\
=&\pm \frac{1}{\sqrt{2}} U_{\alpha\beta}(j) \delta_{s,j_{+}(j_{-})} \delta_{s_{z},m} 
 \left(
 \left\langle j,m_{+};j_{\pm} \right\rangle_{-}
 \left\langle j_{+}(j_{-}),m;j \right\rangle_{+} 
-\left\langle j,m_{-};j_{\pm} \right\rangle_{+}
 \left\langle j_{+}(j_{-}),m;j \right\rangle_{-}
        \right) .
\end{align}
From Eqs.~(\ref{eq:decomposition}) and (\ref{eq:inner_product}) we have
\begin{align} \label{eq:matix_element_negative}
&\hspace*{-2cm}\left\langle \xi^{(l)}(s,s_{z},\beta) \right|
\left(\rho_{-}^{[N]}\right)^{-\frac{1}{2}}
\left| \xi^{(l)}(s^{\prime},s_{z}^{\prime},\beta^{\prime}) \right\rangle \nonumber \\
=&\frac{1}{2} \delta_{s,s^{\prime}} \delta_{s_{z},s_{z}^{\prime}} \delta_{\beta,\beta^{\prime}}
\left( \lambda_{s_{-}}^{-}\right)^{-\frac{1}{2}}
\left| \left\langle s_{-},s_{z+};s \right\rangle_{-}
               \left\langle s,s_{z};s_{-} \right\rangle_{+}
              -\left\langle s_{-},s_{z-};s \right\rangle_{+}
               \left\langle s,s_{z};s_{-} \right\rangle_{-}
        \right| ^{2} \nonumber \\
+&\frac{1}{2} \delta_{s,s^{\prime}} \delta_{s_{z},s_{z}^{\prime}} \delta_{\beta,\beta^{\prime}}
\left( \lambda_{s_{+}}^{-}\right) ^{-\frac{1}{2}} 
\left| \left\langle s_{+},s_{z+};s_{++} \right\rangle_{-}
               \left\langle s,s_{z};s_{+} \right\rangle_{+} 
              -\left\langle s_{+},s_{z-};s_{++} \right\rangle_{+}
               \left\langle s,s_{z};s_{+} \right\rangle_{-}
        \right| ^{2} \nonumber \\
=&\delta_{s,s^{\prime}} \delta_{s_{z},s_{z}^{\prime}} \delta_{\beta,\beta^{\prime}}
\left( \lambda_{s_{-}}^{-}\right) ^{-\frac{1}{2}} \frac{s}{2s+1}
\end{align}
and
\begin{align} \label{eq:matix_element_positive}
&\hspace*{-2cm}\left\langle \xi^{(l)}(s,s_{z},\beta) \right|
\left(\rho_{+}^{[N]}\right)^{-\frac{1}{2}}
\left| \xi^{(l)}(s^{\prime},s_{z}^{\prime},\beta^{\prime})\right\rangle \nonumber \\
=&\frac{1}{2} \delta_{s,s^{\prime}} \delta_{s_{z},s_{z}^{\prime}} \delta_{\beta,\beta^{\prime}}
\left( \lambda_{s_{-}}^{+}\right) ^{-\frac{1}{2}} 
\left| \left\langle s_{-},s_{z+};s_{--} \right\rangle_{-}
               \left\langle s,s_{z};s_{-} \right\rangle_{+}
              -\left\langle s_{-},s_{z-};s_{--} \right\rangle_{+}
               \left\langle s,s_{z};s_{-} \right\rangle_{-}
       \right| ^{2} \nonumber \\
+&\frac{1}{2} \delta_{s,s^{\prime}} \delta_{s_{z},s_{z}^{\prime}} \delta_{\beta,\beta^{\prime}}
\left( \lambda_{s_{+}}^{+}\right) ^{-\frac{1}{2}}
 \left| \left\langle s_{+},s_{z+};s \right\rangle_{-}
               \left\langle s,s_{z};s_{+} \right\rangle_{+}
              -\left\langle s_{+},s_{z-};s \right\rangle_{+}
               \left\langle s,s_{z};s_{+} \right\rangle_{-}
       \right| ^{2} \nonumber \\
=&\delta_{s,s^{\prime}} \delta_{s_{z},s_{z}^{\prime}} \delta_{\beta,\beta^{\prime}}
\left( \lambda_{s+1/2}^{+} \right)^{-\frac{1}{2}} \frac{s+1}{2s+1}
\end{align}
The sum of these two [Eqs.~(\ref{eq:matix_element_negative}) and (\ref{eq:matix_element_positive})] yields 
Eq.~(\ref{eq:matrix_element}) with Eq.~(\ref{eq:c}).

\end{widetext}
%


\begin{thebibliography}{10}
\bibitem{Universal}
Even if $|\varepsilon\rangle$ is generated by applying
$\varepsilon^{\otimes N}$, the scheme still provides us a considerable benefit
such that we can perform $\varepsilon^{\otimes N}$ before getting an input
state (see \protect\cite{Brukner03a}).
\end{thebibliography}
\end{document}